\documentclass[aps,reprint,superscriptaddress,article]{revtex4-2}
\usepackage{bm, amsmath, amsfonts, amssymb, braket,mathtools,upgreek,amsthm}

\usepackage{multirow}
\usepackage{float}
\usepackage{color}
\usepackage{comment}
\usepackage{subfigure}
\usepackage{lipsum}
\usepackage{appendix}
\usepackage{dcolumn}
\usepackage[colorlinks=true, letterpaper=true, pdfstartview=FitV, linkcolor=blue, citecolor=blue, urlcolor=blue]{hyperref}

\begin{document}

\newcommand{\ii}{\text{i}}
\newcommand{\U}{U}
\newcommand{\V}{V}
\newcommand{\BZ}{\left[ 0, 2\pi \right]}
\newcommand{\CZ}{\left[ 0, 1 \right]}

\title{Beyond characteristic equations: A unified one-dimensional non-Bloch band theory via wavefunction data}

\author{Haoshu Li}
\email{lihaoshu@ustc.edu.cn}
\affiliation{Department of Physics, University of Science and Technology of China, Hefei, Anhui 230026, China}

\begin{abstract}
    Non-Hermitian systems play a central role in nonequilibrium physics, where determining the energy spectrum under open boundary conditions is a fundamental problem. Non-Bloch band theory, based on the characteristic equation $\text{det}[E - H(\beta)] = 0$, has emerged as a key tool for this task. However, we show that this framework becomes insufficient in systems with certain symmetries, where identical characteristic equations can yield different spectra. To resolve this, we develop a unified theory that incorporates additional wavefunction information beyond the characteristic equation. Our framework accurately captures spectral properties such as the energy spectrum and the end-to-end signal response in a broad class of systems, particularly those with high symmetry. It reveals the essential role of wavefunction information and symmetry in shaping non-Hermitian band theory.
\end{abstract}

\maketitle

\textcolor{blue}{\emph{Introduction.---}}
In recent years, there has been a remarkable surge in research interest surrounding non-Hermitian systems. Such systems naturally arise in open quantum systems \cite{malzard2015, open1, open3, open4, open5} and in photonic setups with gain and loss \cite{gain1, gain2, gain3, gain4, gain5, gain6, gain7, gain8, gain9, gain10, gain11}. The distinctive spectral characteristics of non-Hermitian systems, particularly the emergence of exceptional points, have been recognized as playing a central role in a wide range of applications including lasing and coherent perfect absorption \cite{doi:10.1073/pnas.1603318113, doi:10.1126/science.aar7709, doi:10.1126/science.abj1028, slhy-f76q, PhysRevLett.106.093902}. Among various non-Hermitian phenomena, the non-Hermitian skin effect (NHSE) \cite{yao2018, yao20182, Thomale2019, londhi2019, song2019, PhysRevResearch.1.023013, origin2020} has attracted particular attention due to its profound implications for fundamental physical principles. In particular, the NHSE leads to a striking breakdown of conventional Bloch band theory and a collapse of the bulk-boundary correspondence, thereby challenging the foundational understanding of band structures in Hermitian systems.

The breakdown of conventional Bloch band theory in predicting open-boundary energy spectra poses a fundamental challenge in non-Hermitian system studies, demanding innovative approaches to compute the energy spectra. The theory that addresses this issue is referred to as non-Bloch band theory \cite{yao2018, PhysRevLett.123.066404, PhysRevB.110.205429, PhysRevX.14.021011}. The main idea can be summarized as follows: replace the conventional Bloch wavevector through the substitution $e^{i k} \rightarrow \beta$, thereby mapping the original non-Hermitian Bloch Hamiltonian $H(k)$ to its non-Bloch counterpart $H(\beta)$. The spectral determination process hinges on solving the characteristic equation $\text{det}[E-H(\beta)]=0$. By solving the zeros of this equation with respect to $\beta$ and applying the condition to get the continuum bands, one can trace out a complex manifold of $\beta$ values, which is referred to as the generalized Brillouin zone (GBZ) \cite{yao2018, yao20182, PhysRevLett.123.066404, PhysRevLett.125.226402, PhysRevB.100.035102, PhysRevLett.124.066602, PhysRevB.101.195147, PhysRevLett.123.246801, PhysRevB.102.085151, PhysRevLett.125.186802, RevModPhys.93.015005, doi:10.1080/00018732.2021.1876991}. This complex-plane trajectory fundamentally serves as the spectral determinant that replaces conventional Brillouin zone concepts. For convenience, we call the condition to get the continuum bands \emph{GBZ condition}. The above description tells that the input information of the non-Bloch theory is the characteristic function $\text{det}[E-H(\beta)]$, and once this is determined, the energy spectrum ``seemingly" can be obtained through a standard procedure.

Although the GBZ condition derived in non-Bloch band theory successfully applies to many systems \cite{PhysRevLett.123.066404}, it has been shown to fail in certain cases where specific symmetries are present \cite{PhysRevB.101.195147, kaneshiro2025, PhysRevB.109.035131, PhysRevB.107.195149}. A common remedy is to modify the GBZ condition in a manner tailored to the corresponding symmetry. However, we find this approach largely \emph{ad hoc} and lacking in generality, which we regard as unsatisfactory. In parallel, other studies have explored this issue in more general contexts based on the algebraic structure of the characteristic function $\text{det}[E-H(\beta)]$ \cite{PhysRevLett.125.226402, PhysRevA.110.012209}. These works suggest that changes in the GBZ condition can be systematically understood through this algebraic structure. In our recent investigation, we present a counterexample that challenges this interpretation: We identify systems that share the same non-Bloch characteristic equation, yet exhibit distinct energy spectra and GBZs. This finding indicates that the characteristic equation alone is insufficient to fully determine the energy spectrum in certain cases, pointing to a fundamental limitation of the current non-Bloch band theory.

This naturally raises the question: What additional information is required, beyond the characteristic function $\text{det}[E - H(\beta)]$, to fully determine the energy spectrum under open boundary conditions? The answer actually lies in the wavefunction data. Here, the wavefunction data mean the following: The roots $\beta$ of the characteristic equation, often referred to as generalized eigenvalues, extend the notion of eigenvalues in linear algebra and encode the spatial decay or growth of the wavefunction. Correspondingly, one can define generalized eigenvectors $v(\beta)$ that satisfy $[E - H(\beta)]v(\beta) = 0$, capturing the internal structure of the wavefunction within a unit cell. We find that the pair $(\beta, v(\beta))$ provides a complete description of the energy spectrum. Within this framework, we achieve a unified formulation of non-Bloch band theory that applies to systems with arbitrary symmetries. Moreover, this approach allows for the straightforward algebraic computation of quantities such as the end-to-end signal amplification factor \cite{PhysRevB.103.L241408, PhysRevB.105.045122, PhysRevX.5.021025, PhysRevLett.122.143901, PhysRevX.3.031001}, which is often difficult to evaluate in systems with high symmetry. For example, in non-Hermitian systems belonging to the symplectic class, the formula for the end-to-end signal amplification factor given in Refs.~\cite{PhysRevB.103.L241408, PhysRevB.105.045122} does not hold. To the best of our knowledge, no analogous formula applicable to such systems has been reported in the literature, and in this work we provide the corresponding expression.

\textcolor{blue}{\emph{Missing information in characteristic function.---}}
\begin{figure*}[htbp]
    \includegraphics[width = 0.95\textwidth]{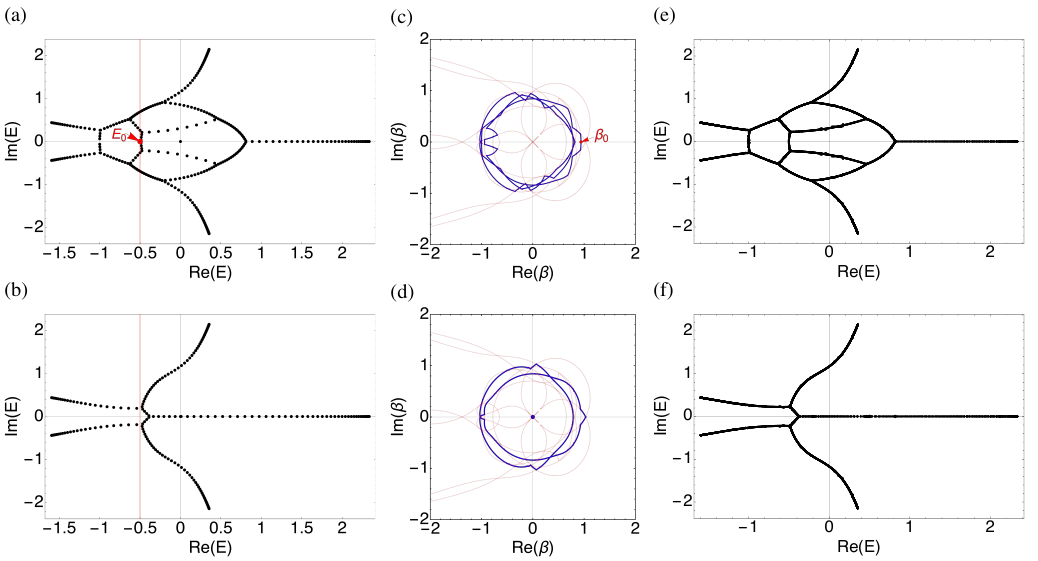}
    \caption[]{The open boundary spectra and GBZs of two models, $H_1(k)$ and $H_2(k)$, which share the same characteristic function. The parameters are set as $t_0=1$ (energy unit) and $a=b=c=d=1$ (dimensionless). Panels (a), (c), and (e) correspond to properties of $H_1$, while panels (b), (d), and (f) correspond to those of $H_2$. (a), (b) Numerically computed spectra for a finite chain of length $L=100$. The energy $E_0 = -1/2$ lies within the spectrum of $H_1$ but not within that of $H_2$. Faint red vertical lines represent Re($E$) = -1/2 in both panels. (c), (d) The GBZs of two models, shown as blue curves. Both models share the same auxiliary GBZ, plotted as faint red curves for reference. Notably, the GBZ in panel (d) includes $\beta=0$. In the GBZ of $H_1$, the mode with generalized momentum $\beta_0$ corresponds to the energy $E_0$, whereas $\beta_0$ does not lie on the GBZ of $H_2$. (e), (f) Spectra reconstructed from the GBZs (c) and (d), respectively, representing the open boundary spectra in the limit $L\rightarrow \infty$.}
    \label{fig: 1}
\end{figure*}
Previous studies have all taken the characteristic function as their starting point, whereas our work is motivated by a counterexample demonstrating that using the characteristic function alone as the input of the theory is fundamentally incomplete. Figure \ref{fig: 1} shows two triple-band non-Hermitian Hamiltonians with the same characteristic function. Figure \ref{fig: 1}(a) is the open boundary spectrum of the Bloch Hamiltonian $H_1(k) = \left(
    \begin{array}{ccc}
     t_0 e^{i k}  & b \, t_0 & 0 \\
     a \, t_0 e^{i k} & t_0 e^{i k} & d \, t_0 e^{-i k} \\
     0 & c \, t_0 e^{-i k} & t_0 e^{i k} \\
    \end{array}
    \right) $ and Fig.~\ref{fig: 1}(b) is the open boundary spectrum of the Bloch Hamiltonian $H_2(k) = \left(
        \begin{array}{ccc}
         t_0 e^{i k} & 0 & 0 \\
         0 & t_0 e^{i k} & t_0 \left(a \, b \, e^{2 i k}+c \, d \, e^{-i k}\right) \\
         0 & t_0 e^{-i k} & t_0 e^{i k} \\
        \end{array}
\right) $. By performing a variable substitution $e^{i k} \rightarrow \beta$, both models result in the same characteristic function $\text{det}[E-H_i(\beta)] = \left(E -t_0 \beta\right) \left(t_0^2 \beta ^2-2 t_0 E \beta-a b t_0^2 \beta +E^2-\frac{c d t_0^2}{\beta ^2}\right), i=1,2$. However, the energy spectra of the two not only are different but also cannot be contained within each other. This difference highlights a fundamental limitation of non-Bloch band theory when the characteristic function is taken as the sole input: The theory is incomplete. One may observe that $H_2$ is a decoupled Hamiltonian, which causes its GBZ condition to differ from the conventional form. While this observation is correct, a more accurate interpretation is that the symmetry of $H_2$ enforces a modification of the GBZ condition. Indeed, it is not only decoupling but also other types of symmetry, such as the symplectic symmetry, that can alter the GBZ condition. 

To explain the spectrum difference between these two models, we need a new theory with other input data beyond the characteristic function that captures the difference of $H_1(k)$ and $H_2(k)$. Before introducing our modified non-Bloch band theory, we first provide a brief review of the conventional formulation and highlight key aspects that are typically overlooked in its derivation.

\textcolor{blue}{\emph{Review of non-Bloch band theory.---}}
Start from a one-dimensional tight-binding model in an open chain. Assume its Bloch Hamiltonian is given by $H(k)=\sum_{n=-m}^m a_n e^{i k n}$ with $a_{-m},a_{-m+1},\cdots,a_{m}$ being $2m+1$ hopping matrices with dimension $l$, and the corresponding tight-binding Hamiltonian is $H = \sum_l \sum_{n=-m}^m a_n c_{l+n}^{\dagger} c_{l}$. The ansatz for the wavefunction is assumed to be $\psi(n) = \beta^{-n} \psi(0)$, where $\psi(n)$ represents the wavefunction amplitude at the $n$-th unit cell and $\beta$ represents the spatial decay rate. Due to the translation symmetry in the bulk, the bulk equation is a characteristic equation 
\begin{align}
    \text{det}[E-H(\beta)] = 0,
    \label{eq: chara}
\end{align}
where $H(\beta)$ is a reformulation of $H(k)$ by performing a variable substitution $e^{i k} \rightarrow \beta$. This is an algebraic equation of both $\beta$ and $E$. For each $E$, there are $2M$ ($M=m l$) roots of $\beta$ by taking into account the zero root and the root at infinity, which can be sorted by their magnitudes: $|\beta_1(E)| \leqslant |\beta_2(E)| \leqslant \cdots \leqslant |\beta_{2M}(E)|$. The boundary condition gives extra equations, which involve the following \cite{PhysRevLett.123.066404}
\begin{align}
    \sum_{P} F_P \prod_{k \in P} (\beta_k)^L = 0,
    \label{eq: boundary_cond}
\end{align}
where the set $P$ is a subset of the mode index set $\{1,2,\cdots,2M\}$ with $M$ elements, $F_P$ is a scalar for each $P$, and $L$ is the length of the one-dimensional chain. To produce the continuum band, the magnitude of the leading-order term and the next-to-leading-order term of Eq.~(\ref{eq: boundary_cond}) must be equal. This requirement leads to the condition $|\beta_{2M} \beta_{2M-1} \cdots \beta_{M+2}\beta_{M+1}|=|\beta_{2M} \beta_{2M-1} \cdots \beta_{M+2} \beta_{M}|$, which is equivalent to $|\beta_{M}|=|\beta_{M+1}|$ \cite{PhysRevLett.123.066404}. This condition is known as the GBZ condition, as it determines the shape of the GBZ.

However, the above derivation overlooks an important subtlety: It does not account for the possibility that the coefficients $F_P$ of terms $\beta_{2M} \beta_{2M-1} \cdots \beta_{M+2}\beta_{M+1}$ and $\beta_{2M} \beta_{2M-1} \cdots \beta_{M+2} \beta_{M}$ in Eq.~(\ref{eq: boundary_cond}) may vanish, i.e., $F_{M+1,M+2,\ldots,2M}=0$ or $F_{M, M+2, \ldots,2M}=0$. In such cases, these terms no longer represent the true leading-order contributions, and consequently, the standard GBZ condition $|\beta_{M}|=|\beta_{M+1}|$ no longer applies.

As we have shown in Fig.~\ref{fig: 1}, changes in the GBZ condition of this kind cannot be inferred from the characteristic function $\text{det}[E-H(\beta)]$.
In other words, the vanishing of the coefficients $F_P$ is not encoded in the characteristic function, and the central contribution of this work is a criterion for determining when $F_P=0$ by using the extra information beyond the characteristic function, which will be introduced in the following.

\textcolor{blue}{\emph{Wavefunctions and the completeness criterion.---}}
In Eq.~(\ref{eq: chara}), we discuss the characteristic equation whose solutions $\beta_i$ ($i=1,\ldots,2M$)  physically represent the spatial decay or growth rates of the eigenmodes. For each mode $i$, the wavefunction inside the unit cell satisfies
\begin{align}
    [E-H(\beta_i)]v_i=0.
\end{align}
In other word, the wavefunction ansatz of the mode $i$ is $\psi(n) = \beta_i^{-n} v_i$. It is quite easy to solve $v_i$ since $[E-H(\beta_i)]$ is a matrix with dimension $l$, which is a small size. For convenience, we call $\beta_i$ the generalized eigenvalues of the characteristic equation and $v_i$ the generalized eigenvector of the generalized eigenvalue $\beta_i$. Similarly, there are also left generalized eigenvectors $u_i$ that satisfy
\begin{align}
    u_i^T[E-H(\beta_i)]=0.
\end{align}

Left and right generalized eigenvectors all encode the wavefunction information. In the following, we explain how to express the GBZ condition in terms of this wavefunction information. First, note that $M=ml$, where $l$ is the dimension of Hamiltonian $H(\beta)$. We can construct supercell right generalized eigenvectors
\begin{align}
    x_i = \begin{pmatrix}
        v_i \\
        \frac{1}{\beta_i} v_i \\
        \vdots \\
        (\frac{1}{\beta_i})^{m-1} v_i
    \end{pmatrix}
\end{align}
and supercell left generalized eigenvectors
\begin{align}
    y_i = \begin{pmatrix}
        u_i \\
        \beta_i u_i \\
        \vdots \\
        \beta_i^{m-1} v_i
    \end{pmatrix}.
\end{align}
It can be seen that they are corresponding wavefunctions inside the supercell with $m$ unit cells, which is quite easy to calculate since $m=M/l$ is quite small. The above form relies on the assumption that all $\beta_i$ are nondegenerate, and the form of the degenerate case can be seen in the Supplemental Material \cite{supplemental}, which is inspired by Ref.~\cite{gohberg2005matrix} and derived independently in Refs.~\cite{PhysRevB.96.195133, Cobanera_2017}. Now we can express the criterion for determining when $F_P=0$ in Eq.~(\ref{eq: boundary_cond}), i.e., 
\begin{align}
    F_P \neq 0 \Longleftrightarrow  \text{det}\left( \sum_{i \in P} x_i y_i^T \right) \neq 0.
    \label{eq: completeness_cond}
\end{align}
There are $M$ terms in $\sum_{i \in P} x_i y_i^T$ and each term $x_i y_i^T$ is a $M \times M$ matrix of rank 1. Hence Eq.~(\ref{eq: completeness_cond}) can be understood as a biorthogonal completeness relation, it tells whether or not these $M$ modes from the subset $P$ can span the whole Hilbert space. Our additional GBZ condition states that for the term $\prod_{i \in P} (\beta_i)^L$ to be the leading term of Eq.~(\ref{eq: boundary_cond}), the corresponding modes from index set $P$ must form a complete basis, which is expressed by the mathematical equation $\text{det}\left( \sum_{i \in P} x_i y_i^T \right) \neq 0$. In other words, the leading term of Eq.~(\ref{eq: boundary_cond}) is the one for which $\text{det}\left( \sum_{i \in P} x_i y_i^T \right) \neq 0$ and the product $\prod_{k \in P} |\beta_k|$ is maximized, which can be mathematically expressed as 
\begin{align}
    P_{\text{max}} = \mathop{\mathrm{arg}\,\max}_{P \in \{i_1,\ldots,i_M | \text{det}\left( \sum_{j=1}^M x_{i_j} y_{i_j}^T \right) \neq 0\}} \prod_{i \in P} |\beta_i|,
    \label{eq: Leading_term}
\end{align}
where $P_{\text{max}}$ denotes the set of mode indices for the leading term, while ``$\mathop{\mathrm{arg}\,\max}$'' represents the standard mathematical operator for identifying indices of maximal elements - a notation widely used in optimization, mathematics, and computer science.
From Eq.~(\ref{eq: Leading_term}), it can be seen that Eq.~(\ref{eq: completeness_cond}) is the key result of this paper, and we put the mathematical derivation in the Supplemental Material \cite{supplemental}. 

\begin{figure}[htbp]
    \includegraphics[width = 1.0\columnwidth]{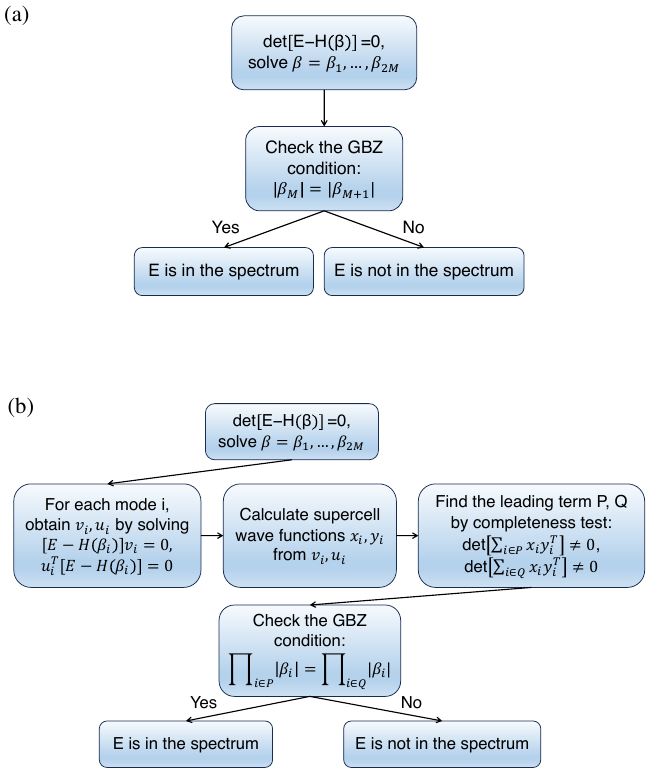}
    \caption[]{(a) Workflow of the original non-Bloch band theory algorithm. Only the most common scenario is shown; in systems with symmetry, the GBZ condition may require case-specific treatment. (b) Workflow of the modified non-Bloch band theory algorithm. The application of this algorithm to specific models $H_1$ and $H_2$ is illustrated in Tables \ref{tab: 1} and \ref{tab: 2}.}
    \label{fig: flowchat}
\end{figure}

Based on Eq.~(\ref{eq: Leading_term}), the algorithm of the modified non-Bloch band theory can be summarized as Fig.~\ref{fig: flowchat}(b) and can be compared to the original non-Bloch band theory shown in Fig.~\ref{fig: flowchat}(a).

The above theoretical framework may seem abstract; we now validate it using the previously discussed triple-band models, as shown in Fig.~\ref{fig: 1}. In Fig.~\ref{fig: 1}(a), we demonstrate that a specific energy level, $E_0 = -1/2$, lies within the spectrum of $H_1$, whereas Fig.~\ref{fig: 1}(b) shows that it does not belong to the spectrum of $H_2$. This discrepancy can be explained using the completeness condition in Eq.~(\ref{eq: completeness_cond}). By substituting $E = -1/2$ into the characteristic equation [Eq.~(\ref{eq: chara})], we obtain six roots: $\beta_1=0,\beta_2=-0.5,\beta_3=-0.94,\beta_4=0.94,\beta_5=-1.06i,\beta_6=1.06i$, ordered by their magnitudes. For each three-element subset $P$ of the index set $\{1,2,\cdots,6\}$, we test the criterion in Eq.~(\ref{eq: completeness_cond}). In Tables \ref{tab: 1} and \ref{tab: 2}, we list several such index subsets $P$ that yield large values of $\prod_{i\in P} |\beta_i|$, corresponding to Hamiltonians $H_1$ and $H_2$, respectively. According to Eq.~(\ref{eq: Leading_term}), the leading-order term and the next-to-leading-order term are indicated in bold. From Table \ref{tab: 1}, it can be seen that $\beta_4 \beta_5 \beta_6$ and $\beta_3 \beta_5 \beta_6$ serve as the leading-order term and next-to-leading-order term, and the GBZ condition $|\beta_4 \beta_5 \beta_6|=|\beta_3 \beta_5 \beta_6|$ is satisfied. This indicates that $E = -1/2$ lies in the spectrum of $H_1$. In contrast, Table \ref{tab: 2} shows that for $H_2$, the leading-order term and next-to-leading-order term are given by $\beta_2 \beta_5 \beta_6$ and $\beta_2 \beta_4 \beta_6$, respectively. However, the GBZ condition $|\beta_2 \beta_5 \beta_6|=|\beta_2 \beta_4 \beta_6|$ is not satisfied, which implies that $E = -1/2$ is not part of the spectrum of $H_2$. Therefore, Eq.~(\ref{eq: completeness_cond}) is crucial to determine the GBZ condition and further the GBZ. Based on this criterion, we compute and plot the GBZs for $H_1$ and $H_2$ in Figs.~\ref{fig: 1}(c) and \ref{fig: 1}(d), respectively. The spectra made from these two GBZs that represent the open boundary spectra at the chain length limit $L\rightarrow \infty$ are shown in Figs.~\ref{fig: 1}(e) and \ref{fig: 1}(f).
\begin{table*}[htbp] \centering
\begin{tabular}{|c|c|c|c|c|c|c|}
    \hline \hline
    $P$ & \textbf{\{4,5,6\}} & \textbf{\{3,5,6\}} & \{3,4,5\} & \{3,4,6\} & \{2,5,6\} & \{2,4,6\} \\
    \hline
    $\prod_{i\in P} |\beta_i|$ & 1.06 & 1.06 & 0.94 & 0.94 & 0.57 & 0.5 \\
    rank$\left( \sum_{i \in P} x_i y_i^T \right)$ & 3 & 3 & 3 & 3 & 3 & 3\\
    det$\left( \sum_{i \in P} x_i y_i^T \right)$ & $\neq 0$ & $\neq 0$ & $\neq 0$ & $\neq 0$ & $\neq 0$ & $\neq 0$ \\
    \hline \hline
\end{tabular}
\caption[]{Biorthogonal completeness of three modes of $H_1$ at energy $E = -1/2$. The index sets $P$ of the leading-order term and the next-to-leading-order term are indicated in bold. The energy $E = -1/2$ lies within the spectrum since the magnitudes of the leading-order term and the next-to-leading-order terms are equal.}
\label{tab: 1}
\end{table*}

\begin{table*}[htbp] \centering
\begin{tabular}{|c|c|c|c|c|c|c|}
    \hline \hline
    $P$ & \{4,5,6\} & \{3,5,6\} & \{3,4,5\} & \{3,4,6\} & \textbf{\{2,5,6\}} & \textbf{\{2,4,6\}} \\
    \hline
    $\prod_{i\in P} |\beta_i|$ & 1.06 & 1.06 & 0.94 & 0.94 & 0.57 & 0.5 \\
    rank$\left( \sum_{i \in P} x_i y_i^T \right)$ & 2 & 2 & 2 & 2 & 3 & 3 \\
    det$\left( \sum_{i \in P} x_i y_i^T \right)$ & $= 0$ & $= 0$ & $= 0$ & $= 0$ & $\neq 0$ & $\neq 0$ \\
    \hline \hline
\end{tabular}
\caption[]{Biorthogonal completeness of three modes of $H_2$ at energy $E = -1/2$. The index sets $P$ of the leading-order term and the next-to-leading-order term are indicated in bold. The energy $E = -1/2$ is not part of the spectrum since the magnitudes of the leading-order term and the next-to-leading-order terms are not equal.}
\label{tab: 2}
\end{table*}

The above discussion shows that the triples $(\beta_i,v_i,u_i)$ can be regarded as the input data of our modifed non-Bloch theory. In mathematical literature, this terminology already exists: The triple $(\beta_i,v_i,u_i)$ is referred to as a \emph{Jordan triple}, and the pair $(\beta_i,v_i)$ as a \emph{Jordan pair} \cite{gohberg2005matrix}.

\textcolor{blue}{\emph{Application in end-to-end signal response.---}} Above we have discussed the importance of the wavefunction information in determining the energy spectrum. We now turn to discuss its role in computing the physical response, i.e., the Green's function. The non-Hermitian Green's function has numerous applications in physics. For instance, it captures the end-to-end response of field coherence in bosonic open systems \cite{PhysRevB.103.L241408}. Even in Hermitian mesoscopic transport problems, it serves as an effective mathematical tool that enables convenient computation of nonreciprocal transport \cite{PhysRevB.107.035306, PhysRevLett.132.156301}. These problems pertain to the end-to-end response of the system. It is known that in general non-Hermitian systems, a signal injected at one end of a chain can be amplified at the other end \cite{PhysRevLett.126.216407}. This amplification follows an exponential law, described by the Green's function as $\|G_{1L}(\omega)\| = \alpha_{\leftarrow}^L$, where $L$ denotes the chain length, $G_{1L}(\omega)=(\omega-H)^{-1}_{1L}$ is the Green's function from the right end to the left end, and $\| \cdot\|$ represents the absolute value in single-band systems and the matrix norm in multiband systems. The exponential factor $\alpha_{\leftarrow}$ is actually $|\beta_M(\omega)|$ (with respect to the convention adopted here) in a general single-band non-Hermitian systems \cite{PhysRevB.103.L241408, PhysRevB.105.045122}. Here, $|\beta_i(\omega)|$ are solutions of $\text{det}[\omega-H(\beta)]=0$ and $|\beta_1(\omega)| \leqslant |\beta_2(\omega)| \leqslant \cdots \leqslant |\beta_{2M}(\omega)|$. Thus, if $|\beta_M(\omega)|>1$, the signal is amplified from right to left. Although single-band systems are well understood, a corresponding result for general multiband systems remains lacking. For example, for a Hamiltonian belonging to the symplectic class, should $\alpha_{\leftarrow}$ be given by $|\beta_M(\omega)|$ or $|\beta_{M+1}(\omega)|$? To address this question, we investigate the multiband non-Hermitian Green's function and derive a universal expression for $\alpha_{\leftarrow}$. The first step is to identify the leading contribution that satisfies the completeness relation in Eq.~(\ref{eq: completeness_cond}); specifically, one must determine an index subset $P_{\text{min}}$ containing $M$ modes that fulfills
\begin{align}
    P_{\text{min}} = \mathop{\mathrm{arg}\,\min}_{P \in \{i_1,\ldots,i_M | \text{det}\left( \sum_{j=1}^M x_{i_j} y_{i_j}^T \right) \neq 0\}} \prod_{i \in P} |\beta_i|.
    \label{eq: P_min}
\end{align}
Similar to Eq.~(\ref{eq: Leading_term}), Eq.~(\ref{eq: P_min}) means that $P_{\text{min}}$ represents the term with nonzero $\text{det}\left( \sum_{i \in P} x_i y_i^T \right)$ and the smallest $\prod_{k \in P} |\beta_k|$.
Then $\alpha_{\leftarrow}$ can be expressed as 
\begin{align}
    \alpha_{\leftarrow} = \mathop{\max}_{i \in P_{\text{min}}} |\beta_i|.
    \label{eq: alpha}
\end{align}
This result admits a straightforward interpretation: The Green's function $G_{1L}(\omega)$ receives contributions from the modes in the index set $P_{\text{min}}$, and can be written as $G_{1L}(\omega) = \sum_{i \in P_{\text{min}}} \beta_i^L G_i(\omega)$. Therefore, in the large $L$ limit, the matrix norm $\|G_{1L}(\omega)\|$ is dominated by the exponential growth associated with the largest $\beta_i$ among these modes, i.e., $\alpha_{\leftarrow} = \mathop{\max}_{i \in P_{\text{min}}} |\beta_i|$.

\begin{figure}[htbp]
    \includegraphics[width = 0.98\columnwidth]{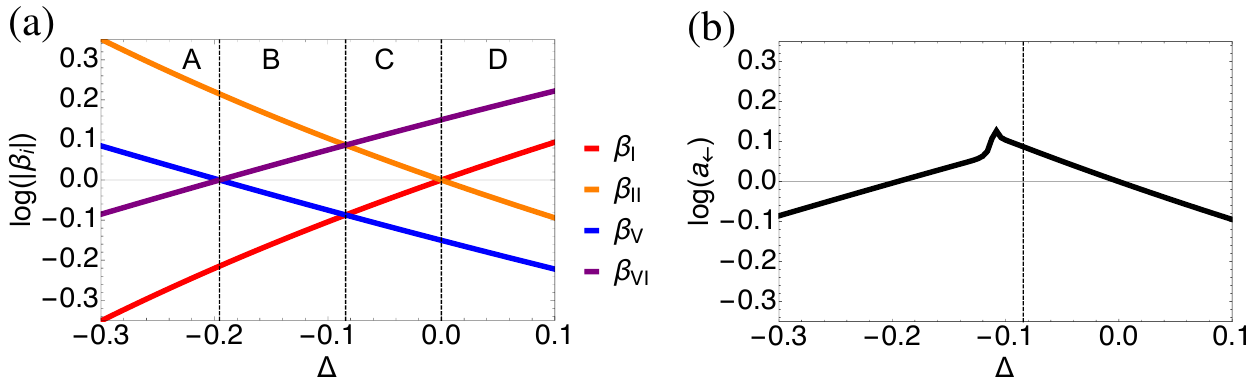}
    \caption[]{Comparison between the solutions of the characteristic equation $\text{det}[\omega-H(\beta)]=0$ and the end-to-end exponential factor $\alpha_{\leftarrow}$. Parameters are set as $t=1.01,u=1,\gamma=1.2$, and $\omega=0.01i$. (a) Magnitudes of analytical solutions $\beta_i$ of the characteristic equation plotted as a function of the parameter $\Delta$. (b) End-to-end exponential factor $\alpha_{\leftarrow}$ as a function of the parameter $\Delta$, obtained via a linear fit of the logarithm of the end-to-end Green's function vs the system length $L$, with $L\in[150,170]$.}
    \label{fig: end2end}
\end{figure}

We test this conclusion using a non-Hermitian Hamiltonian belonging to the symplectic class. The Bloch Hamiltonian is given by $H(k) = \begin{pmatrix}
    0 & D_1(k) \\
    D_2(k) & 0 
\end{pmatrix}$ where $D_1(k) = t \sin k \, \sigma_x + \left( \Delta + u + u \cos k + i \frac{\gamma}{2} \right) \sigma_y$, $D_2(k) = t \sin k \, \sigma_x + \left( \Delta + u + u \cos k \right) \sigma_y$, and $\sigma_i$ are Pauli matrices. This system has a symplectic TRS$^\dagger$ operator $\mathcal{U}_T = -i \sigma_y \otimes \mathbb{I}_{2 \times 2}$, under which the Hamiltonian satisfies $\mathcal{U}_T H(k)^T \mathcal{U}_T^{-1} = H(-k)$. In Fig.~\ref{fig: end2end}, we plot the end-to-end exponential factor $\alpha_{\leftarrow}$ and compare it with the solutions of the characteristic equation $\text{det}[\omega-H(\beta)]=0$. This equation yields eight analytical solutions, which we label using Roman numerals for clarity. Among them, four are shown in Fig.~\ref{fig: end2end}(a); the other four consist of two solutions ($\beta_{\text{IV}},\beta_{\text{VII}}$) that lie above and two ($\beta_{\text{III}},\beta_{\text{VIII}}$) that lie below those depicted. As shown in Fig.~\ref{fig: end2end}(a), there are four crossing points, which are potential transition points of the amplification factor $\alpha_{\leftarrow}$. Based on these crossings, the figure can be divided into four distinct regions, labeled A, B, C, and D. As an illustrative example, we focus on region B. In this region, the completeness relation $\text{det}\left( \sum_{i \in P} x_i y_i^T \right)$ vanishes for the index set $P=\{\text{III},\text{VIII},\text{I},\text{V} \}$, while it remains nonzero for $P=\{\text{III},\text{VIII},\text{I},\text{VI} \}$. These results are obtained from the structure of the wavefunctions (see the Supplemental Material \cite{supplemental}). Therefore, according to the criterion in Eq.~(\ref{eq: P_min}), the minimal index set is $P_{\text{min}}=\{\text{III},\text{VIII},\text{I},\text{VI} \}$, and the corresponding spatial exponential factor $\alpha_{\leftarrow} = |\beta_{\text{VI}}|$.

A universal feature is also observed in Fig.~\ref{fig: end2end}(b): The spatial exponential factor $\alpha_{\leftarrow}$ exhibits a turning point only at the boundary between regions B and C, while the crossing points at $|\beta|=1$ (or $\text{log}(|\beta|)=0$) are not associated with such transitions. This is because the crossing at $|\beta|=1$ violates the completeness condition due to an exchange of wavefunctions, a result that can be rigorously proven in the Supplemental Material \cite{supplemental}. In summary, according to Eqs.~(\ref{eq: P_min}) and (\ref{eq: alpha}), we find that $\alpha_{\leftarrow}=|\beta_{\text{VI}}|$ in regions A and B, and $\alpha_{\leftarrow}=|\beta_{\text{II}}|$ in regions C and D. This prediction is in excellent agreement with the numerical results shown in Fig.~\ref{fig: end2end}(b). However, the question remains: Is $\alpha_{\leftarrow}=|\beta_M(\omega)|$ or $\alpha_{\leftarrow}=|\beta_{M+1}(\omega)|$? Since $|\beta_M(\omega)|$ and $|\beta_{M+1}(\omega)|$ are defined by their magnitudes in ascending order, we can immediately deduce from Fig.~\ref{fig: end2end}(a) that $\alpha_{\leftarrow}=|\beta_M(\omega)|$ in regions A and D and $\alpha_{\leftarrow}=|\beta_{M+1}(\omega)|$ in regions B and C. This stands in stark contrast to the single-band case, where $\alpha_{\leftarrow}$ is always equal to $|\beta_M(\omega)|$. In symplectic class systems, $\alpha_{\leftarrow}$ can take either value, and its precise value is determined by Eqs.~(\ref{eq: P_min}) and (\ref{eq: alpha}). Importantly, distinguishing whether $\alpha_{\leftarrow}=|\beta_M(\omega)|$ or $|\beta_{M+1}(\omega)|$ is crucial for predicting signal amplification, because $|\beta_{M+1}(\omega)|$ that is larger than 1 stands for signal amplification while $|\beta_{M}(\omega)|$ that is smaller than 1 stands for signal decay.

This behavior of the spatial exponential factor $\alpha_{\leftarrow}$ also reveals a deeper connection to the GBZ condition. Specifically, a turning point in $\alpha_{\leftarrow}$ indicates that the resonance frequency $\omega$ enters the spectrum. The universal feature we observed, namely, that crossing points at $|\beta|=1$ do not correspond to turning points, implies that the GBZ condition in the symplectic class cannot take the conventional form $|\beta_M|=|\beta_{M+1}|$. Interestingly, some previous works have proposed an even stronger and more definitive statement that the GBZ condition is given by $|\beta_{M+1}|=|\beta_{M+2}|$. However, we emphasize that the validity of this conclusion is contingent upon a prerequisite: There are no other symmetries that further alter the completeness among the wavefunctions. For example, if the system possesses two distinct symplectic symmetries $\mathcal{U}_{T,1}$ and $\mathcal{U}_{T,2}$, then the spatial exponential factor $\alpha_{\leftarrow}$ may instead correspond to $|\beta_{M+2}(\omega)|$, and the GBZ condition may accordingly shift away from $|\beta_{M+1}|=|\beta_{M+2}|$. Nevertheless, regardless of how many symmetries the system possesses, our expressions for the spatial exponential factor $\alpha_{\leftarrow}$ in Eqs.~(\ref{eq: P_min}) and (\ref{eq: alpha}) remain universally valid. These formulas provide a symmetry-agnostic method for determining $\alpha_{\leftarrow}$, thereby offering a consistent framework even in the presence of complex symmetry structures.

\textcolor{blue}{\emph{Discussion and conclusion.---}} 
The results presented in this work focus on one-dimensional non-Hermitian systems. For systems in higher dimensions, the amoeba formulation has been developed as a nontrivial generalization of one-dimensional non-Bloch band theory \cite{PhysRevX.14.021011}. It is worth noting, however, that the amoeba formulation still relies solely on the characteristic equation as input, and the issue of GBZ condition shifting remains unresolved \cite{kaneshiro2025}. We believe that certain aspects of our results for one-dimensional systems may still be applicable to higher-dimensional cases. For instance, in a two-dimensional system with Hamiltonian $H(\beta_x,\beta_y)$, where $\beta_x$ and $\beta_y$ are the generalized complex momenta, one may fix $\beta_y$ at a value corresponding to a point on the two-dimensional GBZ (i.e., a minimal point in the amoeba) and apply our theory to the resulting pseudo-one-dimensional system along the $x$-direction. In such cases, we expect that the biorthogonal completeness condition given in Eq.~(\ref{eq: completeness_cond}) remains a necessary condition. Whether Eq.~(\ref{eq: completeness_cond}) also serves as a sufficient condition in higher dimensions and how to more effectively treat such systems are important questions that we leave for future study.

The one-dimensional non-Bloch band theory has already been verified in experimental settings \cite{Xiao2020}. However, the nonstandard GBZ conditions in highly symmetric non-Hermitian systems have not yet been experimentally demonstrated. Our modified theory not only provides a non-\emph{ad hoc} framework at the theoretical level, but its generality also makes it well suited for systematic experimental validation of GBZ theory in such symmetric non-Hermitian systems. Experimentally, mechanical oscillators \cite{Li2024, PhysRevLett.131.237201}, photonic crystals \cite{gain1, gain2, gain3}, and electrical circuit \cite{PhysRevB.100.165419, PhysRevB.107.085426} represent promising platforms for such investigations.

In summary, our findings highlight a fundamental limitation of conventional non-Bloch band theory: The characteristic equation of a non-Bloch Hamiltonian alone is insufficient to determine the open boundary spectrum. This realization calls for a significant revision of the theoretical framework, motivating the development of an augmented formulation that systematically incorporates wavefunction information as an essential ingredient in spectral analysis. The resulting generalized theory is broadly applicable across non-Hermitian systems and proves particularly effective in high-symmetry settings, where it bypasses the technical difficulties typically associated with analyzing the GBZ condition and the end-to-end signal response.

{\it Acknowledgement.---}This work was supported by the China Postdoctoral Science Foundation under Grant No. 2023M743398.

\bibliography{Ref}

\end{document}


\newcommand{\ii}{\text{i}}
\newcommand{\U}{U}
\newcommand{\V}{V}
\newcommand{\BZ}{\left[ 0, 2\pi \right]}
\newcommand{\CZ}{\left[ 0, 1 \right]}

\newtheorem{theorem}{Theorem}[section]
\newtheorem{lemma}{Lemma}[section]
\newtheorem{corollary}[theorem]{Corollary}

\title{Supplemental Material for ``Beyond characteristic equations: A unified one-dimensional non-Bloch band theory via wavefunction data''}

\author{Haoshu Li}
\email{lihaoshu@ustc.edu.cn}
\affiliation{Department of Physics, University of Science and Technology of China, Hefei, Anhui 230026, China}

\maketitle
\onecolumngrid
\tableofcontents

\section{Degenerate generalized momentum case}
\label{sec: deg_case}
In the main text, we make an assumption that all generalized momenta are nondegenerate, i.e., $\beta_i \neq \beta_j$ for any $i \neq j$. In this section, we discuss the degenerate case, actually, all procedures are the same except the procedure of constructing supercell wavefunctions. If the wavevector $\beta_i$ is $s_i$-degenerate and is with right wavefunctions $v_{i,1},\ldots,v_{i,s_i}$ (wavefunctions may or may not be the same), then the spatial dependence of supercell right wavefunctions exhibits a power-law prefactor \cite{gohberg2005matrix, PhysRevB.96.195133, Cobanera_2017}, and can be obtained by the following formula \cite{gohberg2005matrix},
\begin{align}
    \begin{pmatrix}
        x_{i,1},\ldots,x_{i,s_i}
    \end{pmatrix} =
    \begin{pmatrix}
        \begin{pmatrix}
            v_{i,1},\ldots,v_{i,s_i}
        \end{pmatrix} \\
        \begin{pmatrix}
            v_{i,1},\ldots,v_{i,s_i}
        \end{pmatrix} J \\
        \vdots \\
        \begin{pmatrix}
            v_{i,1},\ldots,v_{i,s_i}
        \end{pmatrix} J^{m-1}
    \end{pmatrix},
    \label{eq: deg_super}
\end{align}
where $J$ is the Jordan matrix with eigenvalues $\beta_i^{-1}$ and order $s_i$ and $\begin{pmatrix}
    v_{i,1},\ldots,v_{i,s_i}
\end{pmatrix}$ represents a $l \times s_i$ matrix composed by the column vectors $v_{i,1},\ldots,v_{i,s_i}$. For the nondegenerate case, the Jordan matrix becomes a $1 \times 1$ matrix with the single entry $\beta_i^{-1}$, so Eq.~(\ref{eq: deg_super}) reduces to the simple form presented in the main text.

\section{Relation between the left and right wavefunctions}
Actually, the left generalized eigenvectors $u_i$ are rely on the right generalized eigenvectors $v_i$, hence pairs $(\beta_i,v_i)$ nearly encode all information of the system. Without loss of the generality, we give this relation for the system with only the nearest unit cell hopping, i.e., the non-Bloch Hamiltonian of the system is of the form $H(\beta) = h + V \beta + W \beta^{-1}$ where $h,V,W$ are all $M \times M$ matrices. The left generalized eigenvectors $u_i$ can be calculated from the right generalized eigenvectors $v_i$ via the following relation
\begin{align}
    (u_1,\ldots,u_{2M})^T = \begin{pmatrix}
        v_1 & \ldots & v_{2M} \\
        \beta_1^{-1} v_1 & \ldots & \beta_{2M}^{-1} v_{2M}
    \end{pmatrix}^{-1} \begin{pmatrix}
        \mathbb{I}_{M\times 2M} \\
        0_{M\times 2M}
    \end{pmatrix} V^{-1}. 
\end{align}

\section{Analytical derivation of the biorthogonal completeness criterion}
In this section, we give an analytical derivation of the biorthogonal completeness criterion presented in the main text. In this derivation, we use a relation between the Green's function and the energy spectrum, which is that the singular frequency of the Green's function is in the spectrum. Therefore, if the Green's function is of the form $G(\omega) = \frac{A(\omega)}{B(\omega)}$, the singular frequency of the Green's function requires that the denominator satisfies $\text{det}\left[B(\omega)\right]=0$. The equation $\text{det}\left[B(\omega)\right]=0$ naturally corresponds to the boundary condition $\sum_{P} F_P \prod_{k \in P} (\beta_k)^L = 0$ presented in the main text. So the goal of this section is finding the analytical formula of the denominator $B(\omega)$ of the Green's function.

This section is divided into three subsections. Subsection A discusses the explicit form of the denominator $B(\omega)$ of the Green's function and derives the biorthogonal completeness criterion. Subsections B and C present expressions for two types of Green's function. Subsection B addresses the boundary Green's function, while Subsection C discusses the bulk Green's function.

\subsection{Denominator of the Green's function and biorthogonal completeness criterion}
In this section, we give the expression of the Green's function at the boundary by solving the eigenproblem $\left[\omega - H(\beta)\right] v(\beta) = 0$ as introduced in the main text. This method is originally given in Ref.~\cite{PhysRevB.95.235143}, and some changes are made since now the non-Hermitian cases are considered. Our result is an exact formula for the particular matrix block $G_{11}(\omega)$ of the Green's function in the non-Hermitian mutiband system, which is a supplementary result of previous results of Green's function in the non-Hermitian single-band system \cite{PhysRevB.103.L241408, PhysRevB.105.045122}. We divide the whole task in two parts, in this subsection, we obtain the formula of the denominator of the Green's function, which is sufficient to derive our biorthogonal completeness criterion. In the next subsection, we give the complete form of the Green's function at the boundary.

Without loss of generality, we consider the following tight-binding model with only the nearest unit cell hopping (one can always express the Hamiltonian with only the nearest unit cell hopping by extending the unit cell to a larger supercell),
\begin{align}
    H = \sum_{n,\mu,\nu} c^{\dagger}_{n,\mu} h_{\mu\nu} c_{n,\nu} + c^{\dagger}_{n+1,\mu} V_{\mu\nu} c_{n,\nu} + c^{\dagger}_{n,\mu} W_{\mu\nu} c_{n+1,\nu},
    \label{eq: Ham}
\end{align}
where $c^{\dagger}_{n,\mu}$ ($c_{n,\mu}$) is a creation (an annihilation) operator of a particle with index $\mu$ (we assume a unit cell is composed of $M$ degrees of freedom, hence, $\mu=1,\ldots,M$) in the $n$-th unit cell. Hence, the non-Bloch Hamiltonian Hamiltonian of this tight-binding model is $H(\beta) = h+V\beta+W \beta^{-1}$, which is a $M \times M$ matrix.

Consider the following eigenequation in real space
\begin{align}
    (\omega - H) \psi = 0
\end{align}
which can be written in components as 
\begin{align}
    g^{-1}(\omega) \psi(n) - V \psi(n-1) - W \psi(n+1) = \mathbf{0},
\end{align}
where $g^{-1}(\omega) = \omega - h$. Furthermore, we impose the open boundary conditions $\psi(0) = \mathbf{0} = \psi(N+1)$. Assume that $V$ is not singular, the two-components quantities $\Psi(n) = \left[ \psi(n-1)^T, \psi(n)^T \right]^T$ satisfies 
$\Psi(n-1) = T(\omega) \Psi(n)$, where the matrix
\begin{align}
    T(\omega) = \begin{pmatrix}
        V^{-1} g^{-1}(\omega) & -V^{-1} W \\
        \mathbb{I} & 0
    \end{pmatrix}
    \label{eq: left_transfer}
\end{align}
is the transfer matrix.

Denote $G_{11,L}(\omega)$ be the Green's function at the first unit cell in a system with $L$ unit cells, i.e., the Green's boundary at the left boundary. By the Dyson equation, 
\begin{align}
    (g^{-1}(\omega) - W G_{11,L-1}(\omega) V) G_{11,L}(\omega) = \mathbb{I},
    \label{eq: Dyson}
\end{align}
where $g(\omega)=\frac{1}{\omega-h}$ is the Green's function of the unit cell after dropping all inter-cell couplings. Let $X_n(\omega) = G_{11,n}(\omega) V$, Eq.~(\ref{eq: Dyson}) is equivalent to 
\begin{align}
    V^{-1} g^{-1}(\omega) X_L(\omega) - V^{-1} W X_{L-1}(\omega) X_{L}(\omega) = \mathbb{I}
\end{align}
or 
\begin{align}
    \begin{pmatrix}
        \mathbb{I} \\
        X_L(\omega)
    \end{pmatrix} = T(\omega) 
    \begin{pmatrix}
        \mathbb{I} \\
        X_{L-1}(\omega)
    \end{pmatrix} X_L(\omega),
    \label{eq: Dyson_M}
\end{align}
where $T(\omega)$ is the transfer matrix of $\omega-H$ as in Eq.~(\ref{eq: left_transfer}).
Eq.~(\ref{eq: Dyson_M}) can be applied iteratively to obtain 
\begin{align}
    \begin{pmatrix}
        \mathbb{I} \\
        X_L(\omega)
    \end{pmatrix} = T(\omega)^L 
    \begin{pmatrix}
        \mathbb{I} \\
        0
    \end{pmatrix} X_1(\omega) \cdots X_L(\omega),
\end{align}
which implies the final result 
\begin{align}
    G_{11,L}(\omega) = [T(\omega)^L]_{21} [T(\omega)^L]_{11}^{-1} V^{-1},
\end{align}
where $T(\omega)^L$ is divided into four square matrix blocks with the same size and $[T(\omega)^L]_{\mu \nu}$ denote the block of $T(\omega)^L$ at the $\mu$-th row and the $\nu$-th column.
Therefore, in a one-dimensional system with the above Hamiltonian and $L$ unit cells, 
\begin{align}
    G_{11}(\omega) = [T(\omega)^L]_{21} [T(\omega)^L]_{11}^{-1} V^{-1}.
    \label{eq: G_11}
\end{align}
Note that $[T(\omega)^L]_{11}$ is the denominator of the Green's function, whose magnitude is proportional to $\text{det} [T(\omega)^L]_{11}$.

Now we diagonalize $T(\omega)$ to get a simpler form of $[T(\omega)^L]_{11}$. Assume that $T(\omega)$ can be diagonalized as 
\begin{align}
    U^{-1} T(\omega) U = \text{diag}(\beta_1,\beta_2,\ldots,\beta_{2M}),
\end{align}
with $U=(v_1,v_2,\ldots,v_{2M})$ and $U^{-1}=(u_1,u_2,\ldots,u_{2M})^T$, here, $v_i$ are right eigenvectors of $T(\omega)$ and $u_i^T$ are left eigenvectors of $T(\omega)$. Futhermore, by dividing $v_i,u_i$ into two parts with equal length as $v_i = \begin{pmatrix}
    v_{i,1} \\
    v_{i,2}
\end{pmatrix}$ and $u_i = \begin{pmatrix}
    u_{i,1} \\
    u_{i,2}
\end{pmatrix}$, 
\begin{align}
    [T(\omega)^L]_{11} = \sum_{i=1}^{2M} \beta_i^L v_{i,1} u_{i,1}^T.
    \label{eq: TN11}
\end{align}
By Eq.~(\ref{eq: TN11}), $\text{det} [T(\omega)^L]_{11}$ can be expressed as a multivariable polynomial with variables $\beta_1^L,\beta_2^L,
\ldots,\beta_{2M}^L$, i.e., $\text{det} [T(\omega)^L]_{11} = \sum_{i_1,i_2,\ldots,i_M} F_{i_1 i_2 \ldots i_{M}} \beta_{i_1}^L \beta_{i_2}^L \ldots \beta_{i_{M}}^L$. By using the antisymmetric property of the determinant, it can be shown that the coefficient $F_{i_1 i_2 \ldots i_{M}}=0$ if $v_{i_1,1},v_{i_2,1},\ldots,v_{i_{M},1}$ and $u_{i_1,1},u_{i_2,1},\ldots,u_{i_{M},1}$ are linearly dependent respectively, especially, indices $i_1,i_2,\ldots,i_M$ should be distinct. To obtain the coefficient $F_{i_1 i_2 \ldots i_{M}}$, which is called \emph{the mixed discriminant (of subsets)} in mathematics literature, just let $\beta_{i_1}=\beta_{i_2}=\ldots=\beta_{i_{M}}=1$ and all other $\beta_j=0$, the following relation can be derived:
\begin{align}
    F_{i_1 i_2 \ldots i_{M}} & = \left. \left\{\frac{\partial^M}{\partial \beta_{i_1}^L \partial \beta_{i_2}^L \cdots \partial \beta_{i_M}^L} \text{det} [T(\omega)^L]_{11} \right\} \right|_{\beta_1=\beta_2=\cdots=\beta_{2M}=0} \notag \\ 
     & = \text{det} \left( \sum_{\alpha=1}^{M} v_{i_\alpha,1} u_{i_\alpha,1}^T \right).
    \label{eq: coeff_pre}
\end{align}
The final step is expressing $v_{i,1}$ and $u_{i,1}^T$ in terms of supercell wavefunctions. By some algebra, one can show that the right (left) eigenvector of $T(\omega)$ with eigenvalue $\beta_i$ has the form $(x_{i}^T , \beta_i^{-1} x_{i}^T)^T$ [$(y_{i}^T V , -\beta_i^{-1} y_{i}^T W)=(y_{i}^T \beta_i V , - \beta_i y_{i}^T \beta_i^{-1}W)$] with the column vector $x_{i}$ ($y_{i}$) being the right (left) supercell wavefunctions of $\omega - H(\beta_i)$, i.e., $[\omega - H(\beta_i)] x_{i} = 0$ ($y_{i}^T[\omega - H(\beta_i)] = 0$). So we obtain the final result:
\begin{align}
    F_{i_1 i_2 \ldots i_{M}} & \propto \text{det} \left( \sum_{\alpha=1}^{M} x_{i_\alpha} y_{i_\alpha}^T \right) \notag \\
    & \propto \text{det} \left( \sum_{i \in \{i_1,i_2,\ldots,i_{M}\}} x_{i} y_{i}^T \right).
    \label{eq: coeff_final}
\end{align}
Eq.~(\ref{eq: coeff_final}) gives the biorthogonal completeness criterion presented in the main text, i.e., 
\begin{align}
    F_P \neq 0 \Longleftrightarrow  \text{det}\left( \sum_{i \in P} x_i y_i^T \right) \neq 0,
\end{align}
where the set $P$ is a subset of the indices set $\{1,2,\cdots,2M\}$ with $M$ elements.

\subsection{Complete formula of the boundary Green's function}
In this subsection, we continue our derivation of the Green's function. We are interested in the thermodynamics limit $L \rightarrow \infty$, so the leading term of $\sum_{i_1,i_2,\ldots,i_M} F_{i_1 i_2 \ldots i_{2M}} \beta_{i_1}^L \beta_{i_2}^L \ldots \beta_{i_{2M}}^L$ are the most important. The leading term is the term with nonzero coefficient $F_{i_1 i_2 \ldots i_{2M}}$ and the largest $|\beta_{i_1}^L \beta_{i_2}^L \ldots \beta_{i_{2M}}^L|$, i.e., the set of mode indices $P_{\text{max}}$ for the leading term satisfies
\begin{align}
    P_{\text{max}} = \mathop{\mathrm{arg}\,\max}_{P \in \{i_1,\ldots,i_M | \text{det}\left( \sum_{j=1}^M x_{i_j} y_{i_j}^T \right) \neq 0\}} \prod_{i \in P} |\beta_i|,
    \label{eq: Leading_term}
\end{align}
where ``$\mathop{\mathrm{arg}\,\max}$'' represents the standard mathematical operator for identifying indices of maximal elements. We denote the mode indices set $P_{\text{max}}$ of the leading term be $\alpha_1,\alpha_2,\ldots,\alpha_M$. For convenience, we call the generalized momenta of these modes $\beta_{\alpha_1},\beta_{\alpha_2},\ldots,\beta_{\alpha_M}$ \emph{left boundary resonance generalized momenta} or simply \emph{resonance generalized momenta}, since they are the spatial exponential factors of the resonance modes measured at the left boundary at frequency $\omega$.

From the above discussion, eigenvalues of the transfer matrix can be divided into two parts, which are resonance generalized momenta $\beta_{\alpha_1},\beta_{\alpha_2},\ldots,\beta_{\alpha_M}$ and rest eigenvalues $\beta_{\alpha_{M+1}},\beta_{\alpha_{M+2}},\ldots,\beta_{\alpha_{2M}}$. Therefore, according to this grouping, the diagonalization of $T(\omega)$
\begin{align}
    \begin{pmatrix}
        U_{11} & U_{12} \\
        U_{21} & U_{22}
    \end{pmatrix}^{-1} T(\omega) 
    \begin{pmatrix}
        U_{11} & U_{12} \\
        U_{21} & U_{22}
    \end{pmatrix} = 
    \begin{pmatrix}
        B_1 & 0 \\
        0 & B_2
    \end{pmatrix},
\end{align}
where $B_1=\text{diag}(\beta_{\alpha_1},\beta_{\alpha_2},\ldots,\beta_{\alpha_M})$ involoves resonance generalized momenta and $B_2=\text{diag}(\beta_{\alpha_{M+1}},\beta_{\alpha_{M+2}},\ldots,\beta_{\alpha_{2M}})$ involves rest eigenvalues. By Eq.~(\ref{eq: G_11}),
\begin{align}
    G_{11}(\omega) = & \left[ U_{21} B_1^L (U^{-1})_{11} + U_{22} B_2^L (U^{-1})_{21} \right] \notag \\
    & \left[ U_{11} B_1^L (U^{-1})_{11} + U_{12} B_2^L (U^{-1})_{21} \right]^{-1} V^{-1}.
    \label{eq: G11_by_parts}
\end{align}
Due to $U_{11}$ and $(U^{-1})_{11}$ are non-singular (the coefficient $F_{\alpha_1 \alpha_2 \ldots \alpha_{M}}$ of the leading term is nonzero), 
\begin{align}
    G_{11}(\omega) = & \left[ U_{21} + U_{22} B_2^L (U^{-1})_{21} (U^{-1})_{11}^{-1} B_1^{-L} \right] \notag \\
    & \left[ U_{11} + U_{12} B_2^{L} (U^{-1})_{21} (U^{-1})_{11}^{-1} B_1^{-L} \right]^{-1} V^{-1}.
    \label{eq: G11_UB}
\end{align}
Since eigenvalues involved in $B_1$ are larger than those involved in $B_2$, in the thermodynamics limit, 
\begin{align}
    \lim_{L \rightarrow \infty} G_{11}(\omega) = U_{21} U_{11}^{-1} V^{-1}.
    \label{eq: G11_U}
\end{align}
Note that 
\begin{align}
    \begin{pmatrix}
        U_{11} \\
        U_{21} 
    \end{pmatrix} = 
    \begin{pmatrix}
        x_{\alpha_1} , \ldots, x_{\alpha_M} \\
        \beta_{\alpha_1}^{-1} x_{\alpha_1}, \ldots, \beta_{\alpha_M}^{-1} x_{\alpha_M}
    \end{pmatrix}.
\end{align}
Eq.~(\ref{eq: G11_U}) becomes
\begin{align}
    & \lim_{L \rightarrow \infty} G_{11}(\omega) \notag \\
    = & (x_{\alpha_1},\ldots,x_{\alpha_M}) \text{diag}(\beta_{\alpha_1}^{-1},\ldots,\beta_{\alpha_M}^{-1}) (x_{\alpha_1},\ldots,x_{\alpha_M})^{-1} V^{-1},
    \label{eq: G11_therm}
\end{align}
The right hand side of Eq.~(\ref{eq: G11_therm}) is the multiplication of four square matrices $(x_{\alpha_1},\ldots,x_{\alpha_M})$, $\text{diag}(\beta_{\alpha_1}^{-1},\ldots,\beta_{\alpha_M}^{-1})$, $(x_{\alpha_1},\ldots,x_{\alpha_M})^{-1}$ and $V^{-1}$, where $(x_{\alpha_1},\ldots,x_{\alpha_M})$ represents the matrix composed by column vectors $x_{\alpha_1},\ldots,x_{\alpha_M}$. It can be seen from Eq.~(\ref{eq: G11_therm}) that the expression of the left boundary Green's function only involoves the spectral information of $M$ modes with indices $\alpha_1,\alpha_2,\ldots,\alpha_M$. It means that only these $M$ modes resonate with the external input signal of frequency $\omega$ at the left boundary. This also explains why we call $\beta_{\alpha_1},\beta_{\alpha_2},\ldots,\beta_{\alpha_M}$ \emph{left boundary resonance generalized momenta}.

\subsection{Complete formula of the bulk Green's function}
The above is the complete formula of the boundary Green's function, while in this subsection, we give the complete formula of the bulk Green's function. Unlike the previous subsection, our derivation in this one isn't rigorous mathematically. Instead, it emphasizes empirical rules and offers another understanding of the biorthogonal completeness criterion.

Inspired by the formula of the bulk Green's function of single band systems \cite{PhysRevB.103.L241408, PhysRevB.105.045122}, in multiband systems \cite{PhysRevB.107.115412, PhysRevResearch.5.043073}, the Green's function from the unit cell $j$ to the unit cell $j$ should be of the following form:
\begin{align}
    G_{i,j}(\omega) = \int_{C(0, \epsilon)+\sum_{l=1}^M C(\beta_{\alpha_l}, \epsilon)} \frac{\text{d} \beta}{2\pi i \beta} \frac{\beta^{j-i}}{\omega-H(\beta)} ,
    \label{eq: bulk_G}
\end{align}
where $C(\beta_{\alpha_l}, \epsilon)$ is a circle on the complex plane with center $\beta_{\alpha_l}$ and a small enough radius $\epsilon$ such that the circle does not contain other poles. The centers of these circle are $\beta_{\alpha_1},\beta_{\alpha_2},\ldots,\beta_{\alpha_M}$, whose index set $P_{\text{min}}=\{\alpha_1,\ldots,\alpha_M\}$ fulfills
\begin{align}
    P_{\text{min}} = \mathop{\mathrm{arg}\,\min}_{P \in \{i_1,\ldots,i_M | \text{det}\left( \sum_{j=1}^M x_{i_j} y_{i_j}^T \right) \neq 0\}} \prod_{i \in P} |\beta_i|.
    \label{eq: P_min}
\end{align} 

Eq.~(\ref{eq: bulk_G}) can give another understanding of the biorthogonal completeness criterion. To give the biorthogonal completeness criterion, a mathematical result is quite useful \cite{gohberg2005matrix}. First, note that $\omega-H(\beta)$ can be written as $\frac{P_{\omega}(\beta)}{\beta^r}$ where $P_{\omega}(\beta)$ is a matrix polynomial of $\beta$, then the resolvent of $\omega-H(\beta)$ can be expressed as
\begin{align}
    \frac{1}{\omega-H(\beta)} \propto \begin{pmatrix}
        x_1 & x_2 & \ldots & x_{2M} 
    \end{pmatrix}
    \frac{\beta^r}{\beta \mathbb{I}_{2M \times 2M} -J} 
    \begin{pmatrix}
        y_1^T \\ y_2^T \\ \vdots \\ y_{2M}^T 
    \end{pmatrix},
    \label{eq: resolvent_pre}
\end{align}
where $J$ is the Jordan matrix of which the $j$th block has eigenvalue $\beta_j$ and order $s_j$ ($\beta_j$ is $s_j$-degenerate). Here, we just consider the nondegenerate case, in which $J=\text{diag}(\beta_1,\beta_2,\ldots,\beta_{2M})$. So Eq.~(\ref{eq: resolvent_pre}) becomes
\begin{align}
    \frac{1}{\omega-H(\beta)} \propto \sum_{i=1}^{2M} \frac{\beta^r x_i y_i^T}{\beta - \beta_i}.
    \label{eq: resolvent}
\end{align}
Substitute Eq.~(\ref{eq: resolvent}) into Eq.~(\ref{eq: bulk_G}) and apply the Residue theorem, and assume $j-i\geqslant 1-r$ to get rid of the extra singular point $\beta=0$, we can obatin 
\begin{align}
    G_{i,j}(\omega) \propto \sum_{l=1}^{M} \beta_{\alpha_l}^{j-i+r-1} \, x_{\alpha_l} y_{\alpha_l}^T.
\end{align}
For the noninteracting system, the Green's function does not have the zero point, which means that $G_{i,j}(\omega)$ is nonsingular. Therefore, we obtain the biorthogonal completeness criterion again, i.e.,
\begin{align}
    \text{det} \left( \sum_{l=1}^{M} x_{\alpha_l} y_{\alpha_l}^T \right) \neq 0.
\end{align}

Conversely, the importance of this biorthogonal completeness criterion can also be seen. Failure to meet it can lead to the selection of the wrong integration path in Eq.~(\ref{eq: bulk_G}), thereby calculating an incorrect Green's function. 

\section{Symmetry constraints of single band systems}
In this section, we prove that any symmetries in single band systems can not enforce $F_P=0$, hence, the GBZ condition is always the standard one:
\begin{theorem}
    The GBZ condition of any single band system is always $|\beta_M| = |\beta_{M+1}|$, regardless whether or not the system has any symmetries.
\end{theorem}
It is quite direct to obtain this conclusion. Since we are considering the single band system, all wavefunctions become a scalar, i.e., $v_i=u_i=1$. So the supercell wavefunction $x_i=(1,1/\beta_i,\ldots,(1/\beta_i)^{M-1})^T$ and any two distinct supercell wavefunctions $x_i$ and $x_j$ must be linearly independent unless $\beta_i=\beta_j$. However, when $\beta_i=\beta_j$, the degenerate version of supercell wavefunctions must be used instead, and two modes are still linearly independent (see Sec.~\ref{sec: deg_case}). Therefore, any $M$ modes can span the whole Hilbert space and satisfy the completeness criterion $\text{det}\left( \sum_{i \in P} x_i y_i^T \right) \neq 0$, especially modes with spatial decay/growth rates $\beta_{2M},\beta_{2M-1},\ldots,\beta_{M+2},\beta_{M+1}$ and $\beta_{2M},\beta_{2M-1},\ldots,\beta_{M+2},\beta_{M}$. The equality of magnitudes of these two leading term $|\beta_{2M}\beta_{2M-1}\cdots\beta_{M+2}\beta_{M+1}|=|\beta_{2M}\beta_{2M-1}\cdots\beta_{M+2}\beta_{M}|$ implies the GBZ condition $|\beta_M| = |\beta_{M+1}|$.

\section{Symmetry constraints enforced by the symplectic class}
In this section, we derive in detail symmetry constraints of wavefunctions in the symplectic class and reproduce the conclusion about non-Hermitian system in the symplectic class \cite{PhysRevB.101.195147, kaneshiro2025}:
\begin{theorem}
    \label{thm: sym_GBZ}
    The GBZ condition of non-Hermitian Hamiltonian in the symplectic class can not be $|\beta_M| = |\beta_{M+1}|$.
\end{theorem}

In the symplectic class, the non-Bloch Hamiltonian satisfies $\mathcal{U}_T H(1/\beta) \mathcal{U}_T^{-1} = H(\beta)^T$, and $\mathcal{U}_T$ is a unitary symmetry operator satisfies $\mathcal{U}_T \mathcal{U}_T^* = -1$. It implies that solutions of $\text{det}[E-H(\beta)] = 0$ come into pairs $\beta_i \leftrightarrow \beta_{i'}=1/\beta_i$ and $y_{i'}^T x_{i}=0$. We hence make a convention that $|\beta_1| \leqslant |\beta_2| \leqslant \cdots \leqslant |\beta_M| \leqslant |\beta_{M+1}| \leqslant \cdots \leqslant |\beta_{2M}|$ with $\beta_i=1/\beta_{2M-i+1}$. To prove Theorem~\ref{thm: sym_GBZ}, we need the following lemma, which will be used first and proved later. 
\begin{lemma}
    \label{lemma: sym}
    Denote $X_1=\begin{pmatrix}
        x_{2M}, x_{2M-1} , \cdots , x_{M+1}
    \end{pmatrix}$ and $X_2=\begin{pmatrix}
        x_1, x_2 ,\cdots, x_M
    \end{pmatrix}$ both are composited by supercell wavefunctions as column vectors. If $X_1$ is nonsingular, the diagonal elements of $X_1^{-1}X_2$ must be 0.
\end{lemma}

Now we prove Theorem.~\ref{thm: sym_GBZ}: \\
First, we assume $X_1$ is nonsingular, otherwise, it impiles that the wavefunctions group $x_{2M}, x_{2M-1} , \cdots , x_{M+1}$ does not meet the completeness criterion $\text{det}\left( \sum_{i =M+1}^{2M} x_i y_i^T \right) \neq 0$ and $\beta_{2M} \beta_{2M-1}\cdots \beta_{M+1}$ is not the leading term, which breaks the standard GBZ condition $|\beta_M| = |\beta_{M+1}|$. With the assumption that $X_1$ is nonsingular, denote $X_1^{-1}=\begin{pmatrix}
        x_{2M}^{\bot}, x_{2M-1}^{\bot} , \cdots , x_{M+1}^{\bot}
\end{pmatrix}^T$, it follows from $X_1^{-1} X_1=\mathbb{I}$ that $(x_{M+1}^{\bot})^T x_j=0$ for $j=M+2,M+3,\ldots,2M$. Futhermore, by Lemma.~\ref{lemma: sym}, $(x_{M+1}^{\bot})^T x_M=0$ as well. Therefore, the vector space spanned by $x_M,x_{M+2},\ldots,x_{2M}$ ($x_{M+1}$ is substituted by $x_M$) has the codimension $\geqslant 1$ (its biorthogonal complement contains $(x_{M+1}^{\bot})^T$). It means that the wavefunctions group $x_{2M}, x_{2M-1} , \cdots , x_{M}$ does not meet the completeness criterion and $\beta_{2M} \beta_{2M-1}\cdots \beta_{M}$ is not the leading term, which breaks the standard GBZ condition $|\beta_M| = |\beta_{M+1}|$ and we finish the proof.

We can summarize the above proof by one sentence: At least one of $\beta_{2M} \beta_{2M-1}\cdots \beta_{M+1}$ and $\beta_{2M} \beta_{2M-1}\cdots \beta_{M}$ is not the leading term (of course, there is a possibility that neither of them is the leading term) since at least one of the wavefunctions groups $x_{2M}, x_{2M-1} , \cdots , x_{M+1}$ and $x_{2M}, x_{2M-1} , \cdots , x_{M}$ does not meet the completeness criterion.

\subsection{Proof of the Lemma}
In this subsection, we prove Lemma.~\ref{lemma: sym}. The proof itself requires some mathematical techniques and is quite detailed. However, it provides some clues about the characteristics that a type of sysyems with $F_P=0$ need to have. These characteristics do not necessarily require a symplectic symmetry, which offers insights for studying systems with non-standard GBZ condition and beyond the symplectic class.

The first step is to express the Hamiltonian with only the nearest unit cell hopping by extending the unit cell to a larger supercell, and recall the transfer matrix defined before has the form 
\begin{align}
    T(\omega) = \begin{pmatrix}
        V^{-1} g^{-1}(\omega) & -V^{-1} W \\
        \mathbb{I} & 0
    \end{pmatrix}.
\end{align}
The symmetries of the Hamiltonian can be converted to symmetries of the transfer matrix. We can define a symmetry operator as 
\begin{align}
    D = \begin{pmatrix}
        0 & -\mathcal{U}_T W \\
        \mathcal{U}_T V & 0
    \end{pmatrix}.
\end{align}
The symmetries of the Hamiltonian $\mathcal{U}_T^{-1}H^T(\beta)\mathcal{U}_T=H(\frac{1}{\beta})$ and $\mathcal{U}_T^T=-\mathcal{U}_T$ implies that $D^T=D$. Hence, $D$ defines a symmetric bilinear form via $\langle{a},b\rangle{}=a^T D b$. Furthermore, 
\begin{align}
    T^T(\omega) D \, T(\omega) = D,
    \label{eq: orth_trans}
\end{align}
which means that $T(\omega)$ is an orthogonal transformation preserving the symmetric bilinear form $D$.

The transfer matrix $T(\omega)$ can be diagonalized by its eigenvectors:
\begin{align}
    \begin{pmatrix}
        X_1 & X_2 \\
        X_1 B_1^{-1} & X_2 B_2^{-1}
    \end{pmatrix}^{-1} T(\omega)
    \begin{pmatrix}
        X_1 & X_2 \\
        X_1 B_1^{-1} & X_2 B_2^{-1}
    \end{pmatrix} = 
    \begin{pmatrix}
        B_1 & 0 \\
        0 & B_2
    \end{pmatrix},
\end{align}
where $B_1=\mathop{diag}(\beta_{2M},\beta_{2M-1},\ldots,\beta_{M+1})$ and $B_2=\mathop{diag}(\beta_{1},\beta_{2},\ldots,\beta_{M})$ are two diagonal matrices whose diagonals contains eigenvalues of $T(\omega)$. By symmetry relations encoded in Eq.~(\ref{eq: orth_trans}), it follows that 
\begin{align}
    \left[ D \begin{pmatrix}
        X_1 & X_2 \\
        X_1 B_1^{-1} & X_2 B_2^{-1}
    \end{pmatrix} \right]^T T(\omega) \left\{ \left[ D \begin{pmatrix}
        X_1 & X_2 \\
        X_1 B_1^{-1} & X_2 B_2^{-1}
    \end{pmatrix} \right]^T \right\}^{-1} = \begin{pmatrix}
        B_1^{-1} & 0 \\
        0 & B_2^{-1} 
    \end{pmatrix} = \begin{pmatrix}
        B_2 & 0 \\
        0 & B_1
    \end{pmatrix}.
\end{align}
Therefore, column vectors of $\left\{ \left[ D \begin{pmatrix}
        X_1 & X_2 \\
        X_1 B_1^{-1} & X_2 B_2^{-1}
    \end{pmatrix} \right]^T \right\}^{-1}$ are proportional to column vectors of $\begin{pmatrix}
        X_2 & X_1 \\
        X_2 B_2^{-1} & X_1 B_1^{-1}
    \end{pmatrix}$, which can be written as 
    \begin{align}
        \left\{ \left[ D \begin{pmatrix}
        X_1 & X_2 \\
        X_1 B_1^{-1} & X_2 B_2^{-1}
    \end{pmatrix} \right]^T \right\}^{-1} = \begin{pmatrix}
        X_2 & X_1 \\
        X_2 B_2^{-1} & X_1 B_1^{-1}
    \end{pmatrix} \begin{pmatrix}
        \Lambda_1 & 0 \\
        0 & \Lambda_2
    \end{pmatrix},
    \label{eq: pair_relation}
    \end{align}
    where $\Lambda_1$ and $\Lambda_2$ are two diagonal matrices containing the proportionality coefficients. Physically, it is the symmetry relation between each right eigenvector and the left eigenvector of its Kramer pair. Eq.~(\ref{eq: pair_relation}) can be applied twice to obtain $\begin{pmatrix}
        X_1 & X_2 \\
        X_1 B_1^{-1} & X_2 B_2^{-1}
    \end{pmatrix} \begin{pmatrix}
        0 & \Lambda_2 \\
        \Lambda_1 & 0
    \end{pmatrix} = (D^T)^{-1} D \begin{pmatrix}
        X_1 & X_2 \\
        X_1 B_1^{-1} & X_2 B_2^{-1}
    \end{pmatrix} \begin{pmatrix}
        0 & \Lambda_1 \\
        \Lambda_2 & 0
    \end{pmatrix}$. Due to $D^T=D$, it follows that $\Lambda_1=\Lambda_2$, and we denote $\Lambda_1=\Lambda_2\equiv \Lambda$ for convenience.

The next observation is that Eq.~(\ref{eq: pair_relation}) can be written as 
\begin{align}
    \begin{pmatrix}
        X_1 & X_2 \\
        X_1 B_1^{-1} & X_2 B_2^{-1}
    \end{pmatrix}^{-1} & = \begin{pmatrix}
        \Lambda & 0 \\
        0 & \Lambda
    \end{pmatrix} \begin{pmatrix}
        X_2^T & B_2^{-1} X_2^T \\
        X_1^T & B_1^{-1} X_1^T
    \end{pmatrix} D \notag \\
    & = \begin{pmatrix}
        * & \Lambda X_2^T \mathcal{U}_T V \\
        * & \Lambda X_1^T \mathcal{U}_T V
    \end{pmatrix},
\end{align} 
and it gives that (if $X_1$ is nonsingular)
\begin{align}
    \left[ \begin{pmatrix}
        X_1 & X_2 \\
        X_1 B_1^{-1} & X_2 B_2^{-1}
    \end{pmatrix}^{-1} \right]_{12} \left[ \begin{pmatrix}
        X_1 & X_2 \\
        X_1 B_1^{-1} & X_2 B_2^{-1}
    \end{pmatrix}^{-1} \right]_{22}^{-1} = \Lambda (X_1^{-1}X_2)^T \Lambda^{-1}.
    \label{eq: Lemma_relation1}
\end{align}
On the other hand, the inverse of $\begin{pmatrix}
        X_1 & X_2 \\
        X_1 B_1^{-1} & X_2 B_2^{-1}
    \end{pmatrix}$ can be calculated, and if $X_1$ is nonsingular, it can be expressed in the form of 
\begin{align}
    \begin{pmatrix}
        X_1 & X_2 \\
        X_1 B_1^{-1} & X_2 B_2^{-1}
    \end{pmatrix}^{-1} = \begin{pmatrix}
        * & -X_1^{-1}X_2(X_2 B_2 -X_1 B_1 X_1^{-1} X_2)^{-1} \\
        * & (X_2 B_2 -X_1 B_1 X_1^{-1} X_2)^{-1}
    \end{pmatrix},
\end{align}
and it gives that 
\begin{align}
    \left[ \begin{pmatrix}
        X_1 & X_2 \\
        X_1 B_1^{-1} & X_2 B_2^{-1}
    \end{pmatrix}^{-1} \right]_{12} \left[ \begin{pmatrix}
        X_1 & X_2 \\
        X_1 B_1^{-1} & X_2 B_2^{-1}
    \end{pmatrix}^{-1} \right]_{22}^{-1} = -X_1^{-1}X_2.
    \label{eq: Lemma_relation2}
\end{align}
Compare Eq.~(\ref{eq: Lemma_relation1}) with Eq.~(\ref{eq: Lemma_relation2}), it follows that 
\begin{align}
    -X_1^{-1}X_2 = \Lambda (X_1^{-1}X_2)^T \Lambda^{-1},
\end{align}
which implies that the diagonal elements of $X_1^{-1}X_2$ satisfy
\begin{align}
    -(X_1^{-1}X_2)_{ii} & = \Lambda_{ii} (X_1^{-1}X_2)_{ii} (1/\Lambda_{ii}) \notag \\
    \Leftrightarrow (X_1^{-1}X_2)_{ii} & = 0 \quad \text{for any i}.
\end{align}
Therefore, we finish the proof of Lemma.~\ref{lemma: sym}.

\section{Symmetry constraints enforced by nearly decoupled Hamiltonians}
In this section, we demonstrate that in systems with nearly decoupled Hamiltonians, some coefficients $F_P$ in Eq.~(2) of the main text are enforced to vanish. Here, ``nearly decoupled Hamiltonians'' refers to Hamiltonians that can be transformed into a block upper triangular form:
\begin{align}
    H(\beta)= \begin{pmatrix}
        H_1(\beta) & H_{12}(\beta) \\
        0 & H_2(\beta)
    \end{pmatrix}.
\end{align}
In such systems, the Hamiltonian can be approximately decoupled into two subsystems with Hamiltonians $H_1(\beta)$ and $H_2(\beta)$, with unidirectional couplings between them, i.e., there are only couplings from subsystem 2 to subsystem 1. Consequently, the characteristic function becomes reducible and factorizes as $\text{det}[E-H(\beta)]=\text{det}[E-H_1(\beta)]\text{det}[E-H_2(\beta)]$. Without loss of generality, we assume that $H(\beta)$ only includes nearest-unit-cell hoppings so that supercell wavefunctions $x_i,y_i$ are just unit cell wavefunctions $v_i,u_i$. Let $\beta_{1,i}$ be the roots of $\text{det}[E-H_1(\beta)]$, then its corresponding right wavefunctions $x_{1,i}$ resides in the Hilbert space of $H_1(\beta)$. Let $\beta_{2,i}$ be the roots of $\text{det}[E-H_2(\beta)]$, then its corresponding left wavefunctions $y_{2,i}$ resides in the Hilbert space of $H_2(\beta)$. Suppose the dimension of $H_1(\beta)$ is $M_1$ and the dimension of $H_2(\beta)$ is $M_2$, meaning that the Hilbert space of $H_1(\beta)$ can accommodate at most $M_1$ linearly independent modes, while the Hilbert space of $H_2(\beta)$ can accommodate at most $M_2$ linearly independent modes. Therefore, when considering the completeness criterion $\text{det}\left( \sum_{i \in P} x_i y_i^T \right) \neq 0$, if the mode set $P$ contains more than $M_1$ modes as solutions to $\text{det}[E-H_1(\beta)]$, or more than $M_2$ modes as solutions to $\text{det}[E-H_2(\beta)]$, this criterion fails and implies that $F_P=0$.

Moreover, we would like to emphasize that the reducibility of the characteristic function is not the reason why some coefficients $F_P$ vanish; it just so happens that in the nearly decoupled case discussed here, the two coincide. In other words, the reducibility of the characteristic function does not necessarily imply the vanishing of certain $F_P$. This point has already been illustrated in the main text using the example of the Hamiltonian $H_1$ in Fig.~1.

\section{Wavefunctions of the symplectic Hamiltonian example in the main text}
In the main text, we study a Hamiltonian in the symplectic class, which is 
\begin{align}
    H(k) = \begin{pmatrix}
    0 & D_1(k) \\
    D_2(k) & 0 
\end{pmatrix},
\end{align}
where $D_1(k) = t \sin k \, \sigma_x + \left( \Delta + u + u \cos k + i \frac{\gamma}{2} \right) \sigma_y$, $D_2(k) = t \sin k \, \sigma_x + \left( \Delta + u + u \cos k \right) \sigma_y$, and $\sigma_i$ are Pauli matrices. This system has a symplectic TRS$^\dagger$ operator $\mathcal{U}_T = -i \sigma_y \otimes \mathbb{I}_{2 \times 2}$ and the Hamiltonian satisfies $\mathcal{U}_T H(k)^T \mathcal{U}_T^{-1} = H(-k)$. 

Consider the generalized eigenequation $[\omega-H(\beta_i)] x_i = 0$. The solution is 
\begin{align}
    \beta_{\text{I}}(\omega=0) & = -\frac{\Delta +\sqrt{\Delta ^2+t^2+2 \Delta  u}+u}{t+u} , \; x_{\text{I}}(\omega) = (1+O(\omega),0,0,O(\omega))^T , \notag \\
    \beta_{\text{II}}(\omega=0) & = \frac{\Delta -\sqrt{\Delta ^2+t^2+2 \Delta  u}+u}{t-u} , \; x_{\text{II}}(\omega) = (0, 1+O(\omega), O(\omega), 0)^T , \notag \\
    \beta_{\text{III}}(\omega=0) & = \frac{\Delta -\sqrt{\Delta ^2+t^2+2 \Delta  u}+u}{t+u}, \; x_{\text{III}}(\omega) = (1+O(\omega), 0, 0, O(\omega))^T , \notag \\
    \beta_{\text{IV}}(\omega=0) & = \frac{\Delta +\sqrt{\Delta ^2+t^2+2 \Delta  u}+u}{t-u}, \; x_{\text{IV}}(\omega) = (0, 1+O(\omega), O(\omega), 0)^T , \notag \\
    \beta_{\text{V}}(\omega=0) & = \frac{i \gamma +2 \Delta -\sqrt{4 t^2+(2 \Delta +i \gamma ) (i \gamma +2 \Delta +4 u)}+2 u}{2 (t-u)}, 
    x_{\text{V}}(\omega)  = (O(\omega), 0, 0, 1+O(\omega))^T , \notag \\
    \beta_{\text{VI}}(\omega=0) & = -\frac{i \gamma +2 \Delta +\sqrt{4 (t-u) (t+u)+(i \gamma +2 \Delta +2 u)^2}+2 u}{2 (t+u)} , 
    x_{\text{VI}}(\omega)  = (0, O(\omega), 1+O(\omega), 0)^T , \notag \\
    \beta_{\text{VII}}(\omega=0) & = \frac{i \gamma +2 \Delta +\sqrt{4 t^2+(2 \Delta +i \gamma ) (i \gamma +2 \Delta +4 u)}+2 u}{2 (t-u)}, 
    x_{\text{VII}}(\omega)  = (O(\omega), 0, 0, 1+O(\omega))^T , \notag \\
    \beta_{\text{VIII}}(\omega=0) & = \frac{-i \gamma -2 \Delta +\sqrt{4 (t-u) (t+u)+(i \gamma +2 \Delta +2 u)^2}-2 u}{2 (t+u)} , 
    x_{\text{VIII}}(\omega)  = (0, O(\omega), 1+O(\omega), 0)^T.
    \label{eq: beta_y}
\end{align}
It can be seen that, in order to satisfy the completeness criterion presented in the main text, the mode index set $P_{\text{min}}$ must contain a balanced number of even and odd indices, specifically, a total of four indices, with two even and two odd. However, we would like to point out that the model considered here is relatively simple, in the sense that the wavefunctions corresponding to even and odd indices can be grouped into two separate subspaces. As a result, the minimal value of $\text{rank}\left( \sum_{i \in P} x_i y_i^T \right)$ is $M/2$, which is half the dimension of the Hilbert space. In contrast, for generic Hamiltonians in the symplectic class, the wavefunctions do not exhibit this property, and the minimal rank of $\left( \sum_{i \in P} x_i y_i^T \right)$ is $M-1$.

\bibliography{Ref}